\documentclass[preprint,12pt]{elsarticle}

\usepackage{todonotes}
\usepackage{graphicx}
\usepackage{amssymb}
 \usepackage{amsthm}
\usepackage{hyperref}
\biboptions{sort&compress}
\usepackage{amsmath}

\DeclareMathOperator*{\argmin}{arg\,min}
\begin{document}
\begin{frontmatter}
\title{Structural mediation of human brain activity revealed by white-matter interpolation of fMRI}

 \author[a,b,*]{Anjali Tarun}
 \ead{anjali.tarun@epfl.ch}

\author[c]{Hamid Behjat}
\author[d,e]{David Abramian}
\author[a,b]{Dimitri Van De Ville}

 \address[a]{Institute of Bioengineering, \'Ecole Polytechnique F\'ed\'erale de Lausanne, Geneva, 1202, Switzerland}
 \address[b]{Department of Radiology and Medical Informatics, University of Geneva, Geneva, 1202, Switzerland}
 \address[c]{Center for Medical Image Science and Visualization, University of Link\"oping, Link\"oping, 58183, Sweden}
 \address[d]{Department of Biomedical Engineering, University of Link\"oping, Link\"oping, 58183, Sweden}
 \address[e]{Department of Biomedical Engineering, Lund University, Lund, 22100, Sweden}

\begin{abstract}
Anatomy of the human brain constrains the formation of large-scale functional networks. 
Here, given measured brain activity in gray matter, we interpolate these functional signals into the white matter on a structurally-informed high-resolution voxel-level brain grid. The interpolated volumes reflect the underlying anatomical information, revealing white matter structures that mediate functional signal flow between temporally coherent gray matter regions. Functional connectivity analyses of the interpolated volumes reveal an enriched picture of the default mode network (DMN) and its subcomponents, including how white matter bundles support their formation, thus transcending currently known spatial patterns that are limited within the gray matter only. These subcomponents have distinct structure-function patterns, each of which are differentially recruited during tasks, demonstrating plausible structural mechanisms for functional switching between task-positive and -negative components. This work opens new avenues for integration of brain structure and function, and demonstrates how global patterns of activity arise from a collective interplay of signal propagation along different white matter pathways.
\end{abstract}

\end{frontmatter}

\section*{Introduction}
Coordination of distant neuronal populations gives rise to a vast repertoire of functional networks that underpin human brain function. Using functional magnetic resonance imaging (fMRI), temporally coherent activity can be investigated using measures of functional connectivity (FC). On the other hand, the mediation of inter-regional communication by the anatomical scaffold can be conveniently summarized by structural connectivity (SC) extracted from diffusion-weighted MRI (DW-MRI). Over the past decade, several methods have been proposed to bridge the gap between SC and FC. Seminal works have found evidence for strong statistical interdependence between separately defined SC and FC~\cite{Honey2009,Andrews-Hanna2007,Horn2014,Supekar2010}. Limited by the bivariate and summarizing nature of the analysis, the effect is capturing only a general trend of correlation. Following after were studies that specify regions of interests (ROI) as a FC prior for extracting white matter pathways~\cite{Greicius2009,vandenHeuvel2009}, most of which were specifically concentrated to the analysis of the default mode network (DMN), a set of brain regions that are known be more engaged during rest~\cite{Greicius2003}. In contrast to extracting SC from FC priors, a number of studies have attempted to reproduce brain activity from predefined structural connectomes through numerical simulations~\cite{Galan2008,Honey2007,Deco2011,Deco2012,Deco2013,Goni2014}. 

Studies that extract SC from FC or \textit{vice versa} are mostly hypothesis-driven and entail many explicit assumptions. In order to understand how distributed patterns of functional activity arise from a fixed underlying anatomy, a need for research methodologies that are observer-independent and data-driven are of utmost importance. A common approach for data-driven approaches are based on blind-decomposition techniques. By combining diffusion anisotropy (\textit{e.g.}, fractional anisotropy, axial and radial diffusivities) and classical FC to build a joint SC-FC measure~\cite{Sui2013,Sui2015}, structural and functional alterations in healthy and clinical populations are extracted using multimodal canonical correlation analysis and joint independent component analysis (ICA). On the same line, using ICA, the extraction of concurrent white matter (WM) bundles and gray matter (GM) networks from a tractography-based data matrix that records the number of streamlines running from a GM position to a WM location has been proposed~\cite{OMuircheartaigh2018}. Also using tractography, Calamante and colleagues~\cite{Calamante2017} introduced the concept of track-weighted dynamic FC (TW-dFC), where as a first step, WM streamlines from two structurally connected voxel end-points are obtained. Then, these streamlines are weighted according to the FC at their endpoints over a sliding-window that is arbitrarily assigned. TW-dFC then produces a set of four-dimensional volumes showing the averaged FC between corresponding streamline end points onto the WM at the temporal resolution of the window used.  While these studies have made considerable progress in linking structure and function, none of them jointly and simultaneously integrates functional and diffusion data into a \textit{single} integrated framework, which allows the natural emergence of full-brain spatial patterns that covers both the WM and the GM.

Lately, to further transcend our current understanding of structure-function relationships, attention has been set on studying SC and FC through the lens of more technical frameworks borrowed from other research fields, such as propagator-based methods~\cite{Robinson2012, Robinson2016,Atasoy2016}, and control network theoretical tools~\cite{Gu_2015}. Graph signal processing (GSP) for neuroimaging is another emerging field~\cite{Huang2018}, where initially, a graph is defined by identifying regions in GM as nodes, and mapping strength of their SC through WM to edge weights. Functional data are then interpreted as time-dependent graph signals, on which connectome-informed signal processing operations can be performed. This framework incorporates connectivity through WM, but only as SC between pairs of GM regions. 

In this work, we advance the GSP concept by modeling WM explicitly as nodes of the graph for which local connectivity is known from DW-MRI. This allows relating measures of brain activity on the GM with their mediation through WM; i.e., how particular patterns of brain activity jointly rely on an ensemble of WM pathways. This is done by extending the existing approach from region-wise analysis to a whole-brain, voxel-level perspective. We define the blood--oxygenation level dependent (BOLD) time-series from resting-state fMRI as dynamic signals residing on the nodes of a high resolution (i.e., 850,000 voxels) brain graph. By generalizing principles of classical signal processing in regular domains~\cite{Buades2005,Rudin1992,Chambolle2004,Candes2006} to irregular graphs~\cite{Shuman2013,Chen2014}, we interpolate functional signals, measured on the GM, into the WM, using a whole-brain voxel-wise connectome to guide the process. The recovered volumes provide a novel perspective on how coherent GM regions communicate through various combinations of WM pathways. The application of conventional and dynamic FC tools on the recovered volumes reveals new structure-function relationships of DMN-related networks, illustrating the structural mechanism for the dynamic switching between task-positive and -negative subsystems of the DMN in different phases of working memory and relational task paradigms.

\section*{Results}
\subsection*{Recovery of activity in white matter that is compatible with structural connectivity} 
Signal recovery is a classical task that relates to image denoising~\cite{Buades2005}, signal inpainting~\cite{Rudin1992,Chambolle2004}, and sensing~\cite{Candes2006}. Here we embed the three-dimensional voxel-level grid into a graph where nodes are voxels, and edges of the 26-neighborhood of each node encode strength of structural connectivity as measured by DW-MRI. Functional volumes are preprocessed and upsampled to the same resolution as this structural grid. Nodes in the GM are assigned the values from these functional volumes, while the remaining nodes, in particular, those in the WM, remain unassigned. The values of the unassigned nodes are then recovered by solving an optimization problem that relies on two assumptions: (i) values in the GM nodes should strongly influence the data fitness (and thus remain reasonably unchanged); (ii) whole-brain signals should maintain smoothness according to the graph structure. The assumption of smoothness with respect to the graph ensures that signals are interpolated according to the brain's structural scaffold. The overall pipeline is illustrated in Fig.~\ref{fig1}(A) and explained in details in the Supplementary Information. 
\begin{figure}
\centering
\includegraphics[width=\textwidth]{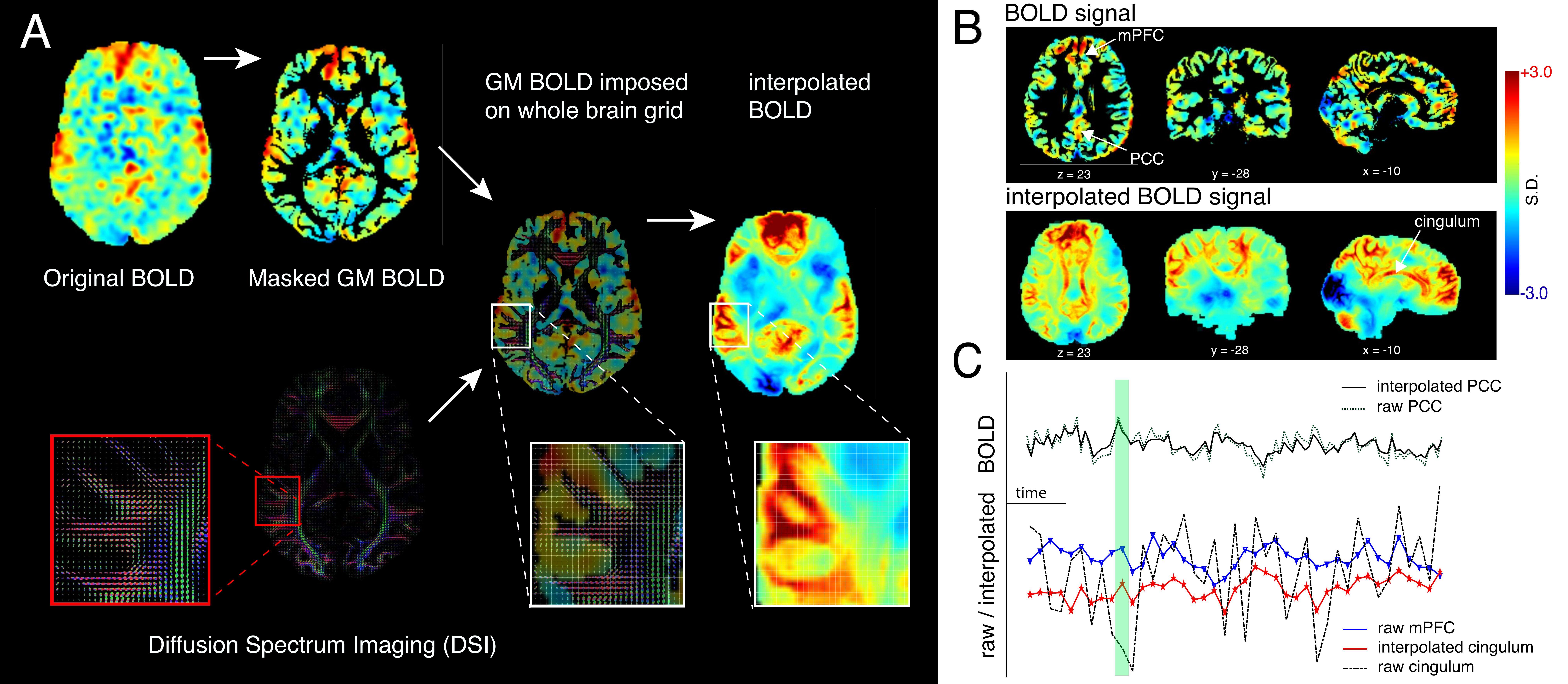}
\caption{\textbf{Workflow of the graph signal recovery framework.} (A) GM BOLD signals are extracted from fMRI volumes, one signal per time instance, through masking the volume by the thresholded probabilistic GM map (threshold=0.3). ODFs associated to all voxels are extracted from the diffusion MRI data. The ODFs are then embedded on a three-dimensional, 26-neighborhood connectivity grid, forming a probabilistic connectome at voxel-level resolution. A signal value can be associated to each voxel on the grid. To initiate signal recovery at each time instance, the associated GM BOLD signal is mapped to the GM voxels within the grid, and the remaining voxels are set to zero. The interpolated BOLD volume is obtained through minimizing a cost function that balances between (1) retaining the GM signal fixed and (2) obtaining a smooth signal over the entire brain grid relative to the underlying structure. (B) Axial, sagittal, and coronal views of a single slice of a BOLD volume is compared to its corresponding interpolated BOLD volume. (C) Original and interpolated BOLD signal time-series of PCC, mPFC and cingulum; the selected time frame in (B) corresponds to the highlighted time point.}
\label{fig1}
\end{figure}
For an illustrative frame of resting-state fMRI, we observe strong activity in the posterior cingulate cortex (PCC) and medial prefrontal cortex (mPFC) for the original and interpolated volumes (Fig.~\ref{fig1} (B)). The latter reveals additional WM signals showing high activation in the cingulum bundle (R/Lcing), atlased using NeuroImaging Tools and Resources Collaboratory (NITRC) repository. In addition to major WM bundles that are clearly captured, we also find short-range WM connectivity within the cortical foldings of the GM~\cite{Koch2002}. The original PCC-averaged signal (Fig.~\ref{fig1}(C)) shows close similarity to its interpolated counterpart, a direct consequence of the second constraint imposed in the signal recovery formulation (\textit{i.e.}, retaining the signals within the GM nodes fixed). In contrast, we find a huge difference in the mean signals within the R/Lcing. PCC, mPFC and recovered R/Lcing signals all show similar trends, while the original R/Lcing signal behaves much more randomly. This observation demonstrates that interpolated functional signals are smooth over the structural brain grid, reflecting in this particular case the existing anatomical pathway (R/Lcing) running from the rostral aspect of the PCC towards the mPFC~\cite{Greicius2009,vandenHeuvel2009}.

\subsection*{PCC-seed connectivity map}
\begin{figure}
\centering
\includegraphics[width = \columnwidth]{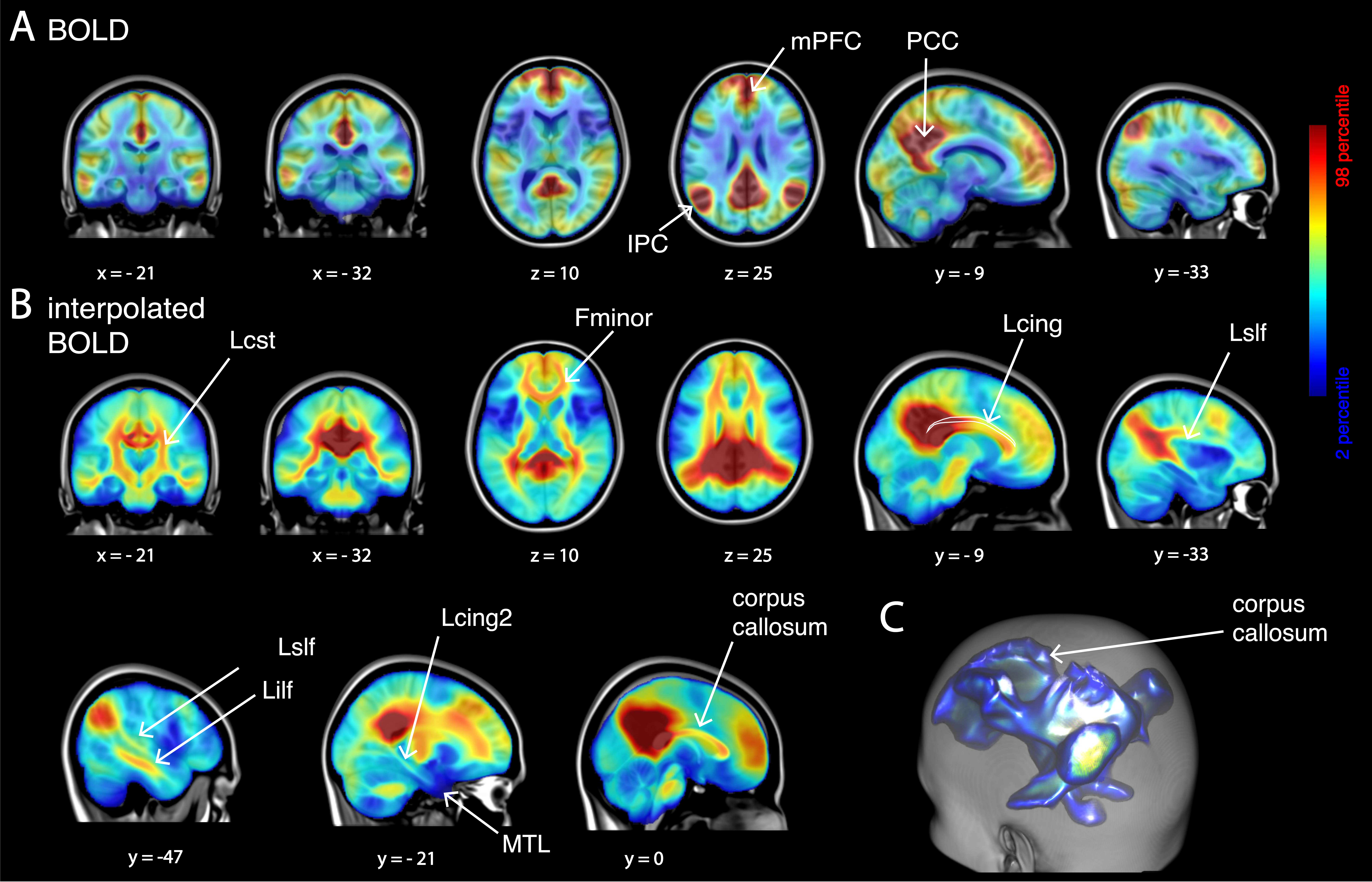}
\caption{\textbf{PCC-seed correlation} for the (A) original BOLD volumes and the (B) interpolated volumes, averaged across all subjects. WM bundles connecting temporally-coherent GM regions are uncovered from interpolated volumes. (C) 3D view of the observed WM structure.}
\label{fig2}
\end{figure}

One conventional method to explore FC is seed-correlation analysis, where the BOLD time course of an ROI is correlated to the activity profiles of all brain voxels to locate temporally coherent areas. In order to capture the joint structural-functional connection of the DMN, we perform a PCC seed-connectivity analysis~\cite{Biswal1995} on the interpolated volumes derived from four sessions of resting-state scans in a population of 51 subjects (244,800 volumes in total). The superior axial slice ($z$=25) in Fig.~\ref{fig2}(A) and (B) shows the PCC and the mPFC, including the bilateral inferior parietal cortex (IPC). In addition to the expected GM FC pattern, Fig.~\ref{fig2}(B) also shows the distinct WM structures that support the structural connections between these temporally coherent cortical regions. The most prominent connections are the R/Lcing that connect the PCC and the mPFC, the forceps minor (Fminor) that provides intra-connectivity between gyri within the mPFC, and the left and right superior longitudinal fasciculi (R/Lslf) that support the long-range connection between the IPC, and the posterior and frontal regions. Interestingly, we also find the two-sided corticospinal tracts (R/Lcst) that traverse the brainstem all the way to the motor cortex, the genu of the corpus callosum connecting the two hemispheres, as well as the bilateral hippocampal cingulum (R/Lcing2) that exit the caudal aspect of the PCC and continue towards the medial temporal lobe (MTL). 
 
\begin{figure}
\centering
\includegraphics[width= \columnwidth]{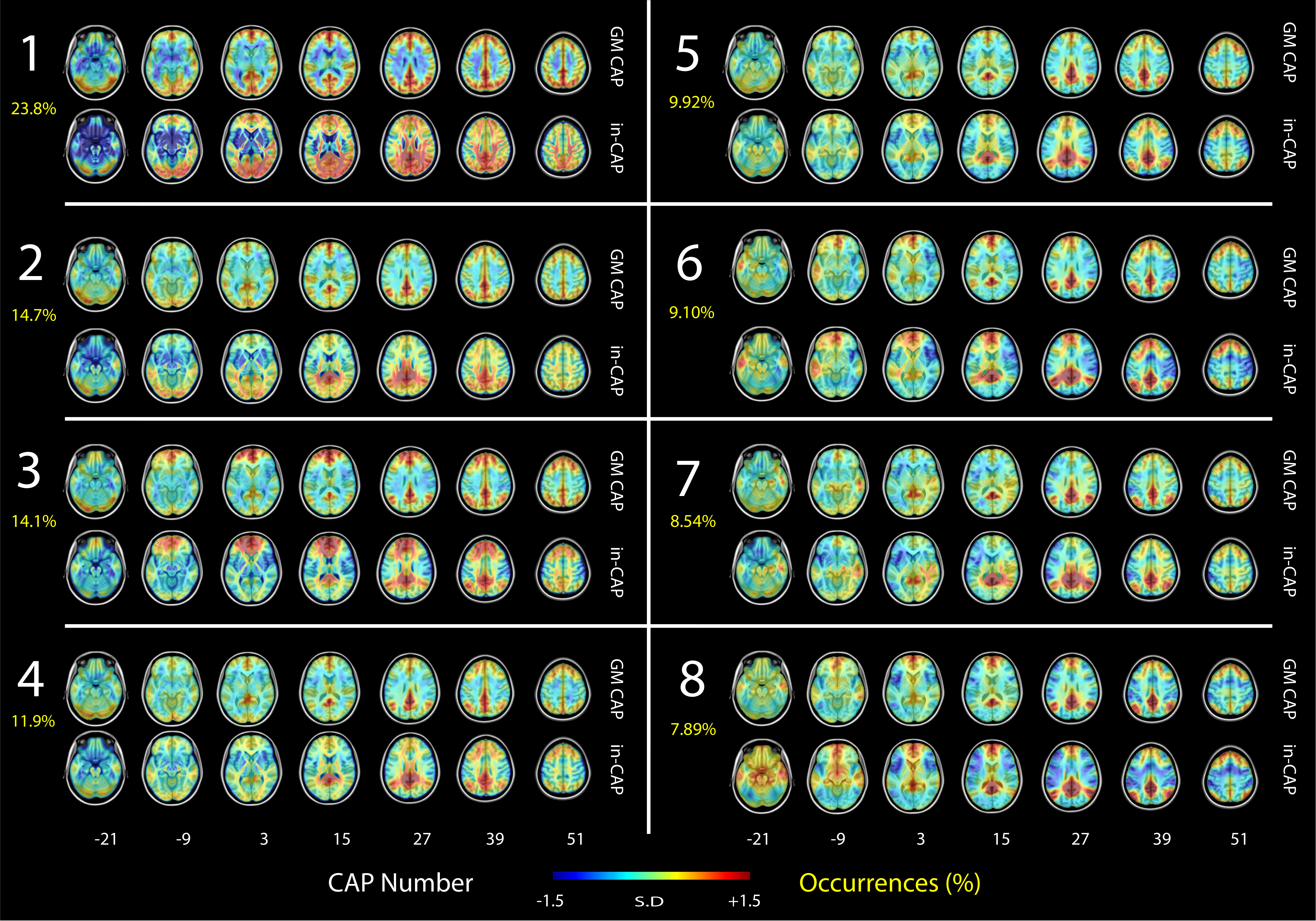}
\caption{\textbf{PCC-seed co-activation patterns (CAPs).} Eight co-activation maps obtained by temporal decomposition of the top 15\% significant frames related to the posterior cingulate cortex (PCC), extracted from (1) the original BOLD volumes (GM CAPs) and (2) interpolated volumes (in-CAPs). The in-CAPs are numbered according to their frequency of occurrence. Visual inspection of the in-CAPs reveals strong GM similarity with the observed GM CAPs. We found eight distinguishable structure-function networks related to the default mode network (DMN), each of which varies in terms of the observed WM structures that conjoin distinct PCC-coherent GM regions.}
\label{fig3}
\end{figure}

\subsection*{Structural mediation of dynamic functional connectivity}
The volume shown in Fig.~\ref{fig1}(B) corresponds to a time point with significant activity in the PCC (green shade, Fig.~\ref{fig1}(C)). FC obtained from conventional seed-correlation analysis can be approximated by averaging all frames for which the seed has high activity~\cite{Liu2013}. This congregation of significantly active frames can also be temporally clustered into distinct and meaningful co-activation patterns (CAPs). Using this approach, we decomposed the DMN pattern into eight PCC-related co-activation patterns (CAPs). Fig.~\ref{fig3} presents the CAPs obtained from the interpolated volumes, termed \textit{interpolated CAPs} (in-CAPs), as well as the CAPs generated from the original BOLD signal, denoted GM CAPs. The resulting spatial patterns reveal a set of WM structures unique to each in-CAP. In-CAP 1 is the most frequently occurring and is characterized by many negative ventral regions, in contrast with the highly positive occipital and frontoparietal regions. In-CAP 2 shows similar characteristics, but with much less pronounced activation in the frontal region, as also seen in its weaker Fminor and R/Lcing. In-CAP 3 exhibits very strong activation in the frontal lobe and negative signals in motor and sensory areas. In-CAPs 4, 5, 6 and 7 all bear high resemblance with the DMN, but vary in their strength of activity in the occipital and frontoparietal regions; these variations are consequently also reflected in Fminor, Fmajor, R/Lcing, R/Lslf, R/Lilf, as well as the bilateral fronto-occipital fasciculus (R/Lifo). In-CAPs 6 and 7 contain highly negative right and left frontoparietal signal, respectively, demonstrating strong hemispherical asymmetry in their activation patterns. Finally, in-CAP 8 is characterized by positive activation within the ventral brain, specifically in the MTL, but highly negative frontoparietal and somatosensory signals---the implicated WM structures are less pronounced in this in-CAP.

\subsection*{Relevance of WM interpolated patterns during rest and task}
Whereas GM CAPs reveal instantaneous co-activation of multiple brain regions, in-CAPs complete the structure-function picture through the addition of distinct and functionally relevant WM structures that interpose spatially-distant GM co-activations. The concurrence of the two CAP types is confirmed by their generally matching frame assignments (Supplementary Fig.~S1). This strong temporal correspondence illustrates the ability of the proposed method in successfully capturing distinct WM structures connecting temporally varying PCC-related networks, without compromising information from the original data. To further study in-CAPs and their correspondence with the DMN subsystems, we computed the average signal within major WM tracks (Fig.~\ref{fig4}(A)) using the NITRC WM atlas, and compared them to the main components of the DMN (Fig.~\ref{fig4}(B)), as classified by Andrews-Hannah and colleagues~\cite{Andrews-Hanna2010}. We observed good correspondence in the spatial patterns of GM regions and WM tracks, especially in terms of activation polarity from the dorsal (positive-valued) to the ventral (negative-valued) side. In-CAP 8, however, is positive across all DMN sub-components, and in contrast to all other in-CAPs, also shows a positive bilateral uncinate fasciculus (L/Runc).

Previous work have found that activations in both the hippocampus and ventral mPFC during retrieval-mediated learning support novel inference~\cite{Zeithamova2012}. We hypothesize that the positive activity of L/Runc in in-CAP 8 reflects its potential role in mediating the coherent activity between the positive temporal poles, hippocampus, and ventral mPFC. To verify this association and to further explore the functional meaning of the obtained in-CAPs, we computed their occurrence probability upon working memory and relational tasks (Fig.~\ref{fig4}(C)). To do so, we interpolated signals into the WM as before, extracted frames with high activation in the PCC, and assigned them to their closest in-CAP. We found that although in-CAP 1 is the most recruited at rest, it mostly occurs during task blocks (as opposed to rest intervals). In contrast, in-CAP 8 occurs the least during resting state, but it is vividly recruited during the rest epochs of the task paradigms. These results are evaluated using a two-factor ANOVA on the occurrences of the different interpolated CAPs during rest and tasks. Results showed highly significant main effects of CAP type and task, as well as the interaction between both, with all corresponding p-values less than $0.0001$.

\section*{Discussion}
\subsection*{General findings}
We modeled the brain grid, together with local, voxel-to-voxel probabilistic connections based on DW-MRI, as a large graph onto which brain activity in GM is interpolated into WM. The interpolated brain volumes remarkably reflected WM structures that support distributed patterns of activity in GM. We explored the neuroscientific relevance of the interpolated volumes by applying conventional and dynamic FC tools, namely seed-correlation~\cite{Biswal1995} and CAP analysis~\cite{Liu2013}, using the PCC as a seed. The resulting in-CAPs unraveled the exquisite complexity of recruited WM patterns underlying typically observed GM co-activations (Fig.~\ref{fig3}). In-CAPs also showed strong neurophysiological relevance, as they revealed the segregation of the DMN into task-positive and task-negative sub-components (Fig.~\ref{fig4}C).

\subsection*{Structure-function relationships of PCC-related networks}
\begin{figure}
\centering
\includegraphics[width=\columnwidth]{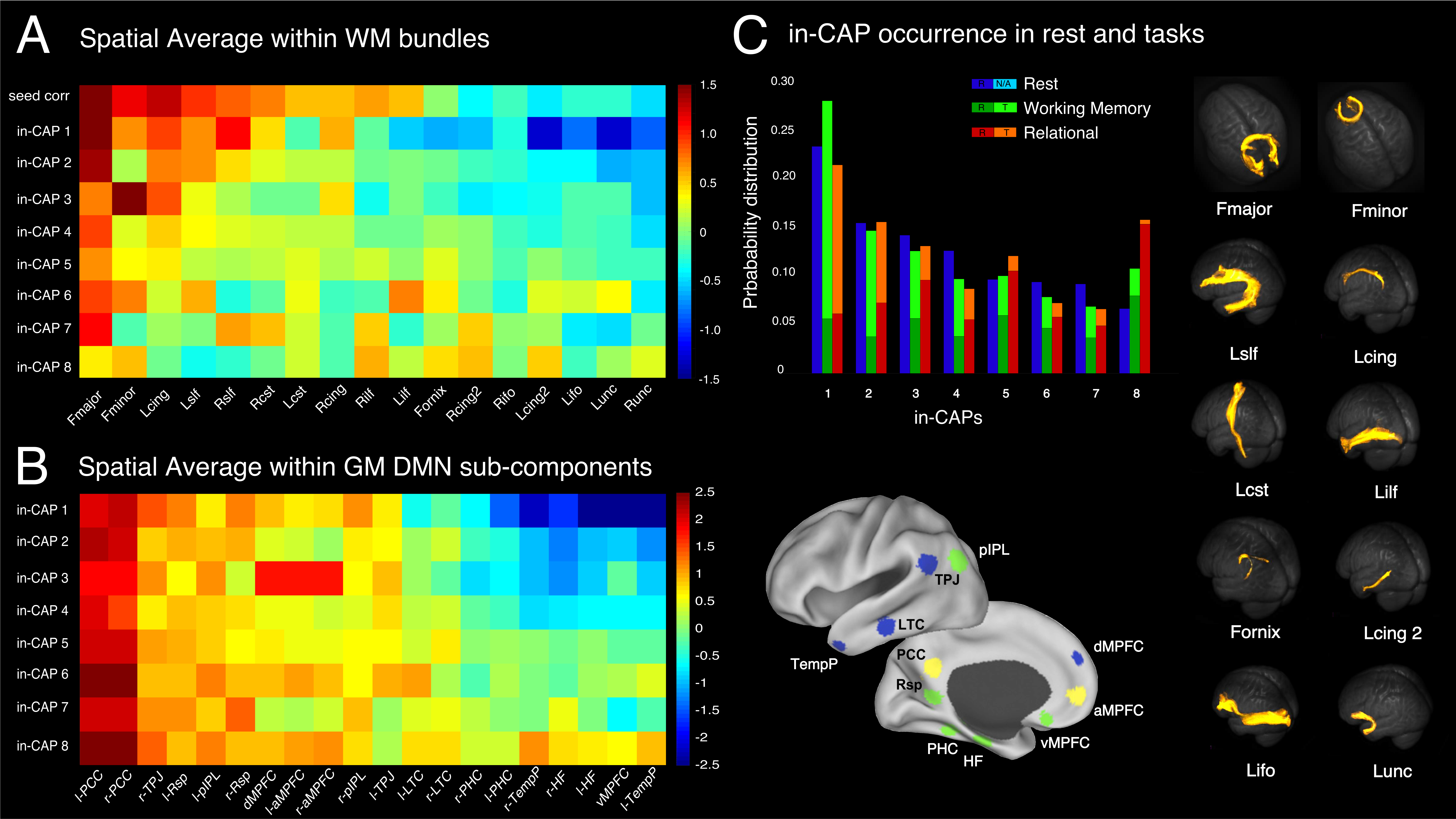}
\caption{\textbf{Spatial and temporal characteristics of in-CAPs.} The overall spatial pattern of the in-CAPs is explored through spatial averaging of signals within (A) 17 major WM bundles, and (B) GM sub-components of the DMN~\cite{Andrews-Hanna2010}. The arrangement of WM structures and DMN sub-components based on their polarity show a concordant pattern of positive-to-negative transition from dorsal-to-ventral regions. (C) The probability of occurrence of in-CAPs in resting-state, working memory, and relational tasks. Each bar in the plot corresponds to the probability of that particular in-CAP to appear in the top 15\% significant volumes of rest and task scans. The stack of colors denotes the proportion of frames that occur during blocks of rest (R) when there is no stimulus presented, and during blocks of task (T) in which subjects are expected to perform memory or relational test.}
\label{fig4}
\end{figure}

Interpolating functional signals measured on GM into the WM has revealed new structural-functional relationships that are akin to the DMN. Fig.~\ref{fig4}(A) shows the signal average within WM bundles observed in our PCC seed-correlation map, consistent with the visually observed tracks in Fig.~\ref{fig2}. Prominent tracks include the R/Lcing, the R/Lslf, and the Fminor in close agreement with previous findings of WM structures associated with the DMN~\cite{Greicius2009, vandenHeuvel2009}. Although seed-correlation analysis is a straightforward measure of FC, it is debased by its transitive nature. The connectivity obtained from this static approach, therefore, summarizes all major WM connections linked to the PCC. On the other hand, CAP analysis can account for time-dependent behavior; thus, the in-CAPs reveal different sets of WM bundles that are characteristic of each of the observed GM CAPs at different timepoints. 

While it has been established that the DMN anatomically consists of the anterior and posterior midline, the bilateral parietal cortex, prefrontal cortex, and the MTL~\cite{Buckner2008}, these regions have been found to functionally dissociate according to the ongoing internal processes~\cite{Andrews-Hanna2010}. We surmise that this functional disintegration is captured by in-CAPs, given that distinct in-CAPs occur in different phases of working and relational memory tasks (Fig.~\ref{fig4}(C)). The MTL activates when internal decisions involve constructing a mental scene based on memory~\cite{Andrews-Hanna2010}. However, despite being dominantly active in this region, in-CAP 8 appears to be more active during blocks of rest, and instead, in-CAPs 1 and 2 dominate in periods of task blocks when the stimulus is presented. These findings suggest that in-CAP 8 is a characteristic \textit{task-negative} structure-function network that activates during the maintenance phase (\textit{i.e.}, when subjects are focused on consolidating an observed stimulus), and conversely, in-CAPs 1 and 2 are \textit{task-positive} structure-function networks that are engaged during the encoding and retrieval periods when the external stimulus is presented. Altogether, our results provide new insights into the well-established dichotomy of the human brain~\cite{Fox2005,Greicius2004}, and clearly demonstrate that the DMN and task-positive networks (\textit{e.g.}, attention network, working memory network) are not exclusively expressing opposite activity~\cite{Piccoli2015,Karahanoglu2015}.

Arranging the WM tracks and the DMN sub-components according to the polarity of their activity in each of the in-CAPs (Fig.~\ref{fig4}(A-B)) reveals a general trend of positive dorsal areas, such as those observed from Fmajor to Lilf, and from left PCC (l-PCC) to left temporal parietal junction (l-TPJ). On the other hand, ventral components (e.g., Rcing2 to Runc, left lateral temporal cortex (l-LTC) to left temporal pole (l-Temp)) are more varied, showing a subtle divide between more negative ventral (in-CAPs 1-4) versus more positive ventral default networks (in-CAPs 5-8). Sagittal slices (Supplementary Fig.~S3) show strong activation in the ventral PCC of in-CAP 8, together with the R/Lcing2, which runs from the caudal aspect of the retrosplenial cortex (RSP) to the hippocampal formation (HF). The bilateral HF, PCC, R/Lcing2, and temporal poles (R/LTempP) also exhibit more positive signal and, therefore, more homogeneous averaged activity. Overall, the above observations highlight the distinctive role of dorsal and ventral parts of the PCC in cognitive control~\cite{Leech2011,Karahanoglu2015}. 
Furthermore, we verify our hypothesis that R/Lunc mediates the coherent activity between the hippocampus and prefrontal cortex in in-CAP 8 during memory consolidation~\cite{Squire2005}. Fittingly, it has been found that the deterioration of R/Lunc lowers working-relational memory performance~\cite{Hanlon2012}.

\subsection*{Methodological perspectives}
The proposed methodology is one of the very few studies that integrates and jointly models functional and diffusion data in an observer-independent and fully data-driven manner. Previous works have relied on blind-decomposition techniques and the use of either diffusion anisotropy measures or tractography algorithms to approximate SC~\cite{Sui2013,Sui2015,OMuircheartaigh2018}. However, ICA's principal assumption of spatial independence leads to components that have low spatial similarity, while measures of diffusion anisotropy are unable to resolve crossing fibers as they are extracted from low-order diffusion tensor models, making their interpretation misleading and prone to erroneous conclusions~\cite{Wheeler-Kingshott2009}. Additionally, the high susceptibility of tractography algorithms in generating false positive~\cite{Maier-Hein2017} suggests cautious to be taken in interpreting tractography-based results. In contrast, our proposed method does not rely on tractography thereby, not only bypassing the complicated task of parameter optimization for extracting WM tracks but also not requiring a parcellation scheme to define ROIs. Our approach keeps the analysis close to the DW-MRI data by working directly with local reconstructions of orientation distribution functions, which are consequently encoded, both in direction and magnitude, to build the brain graph. Furthermore, it is worth noting that despite the superficial similarity, there are fundamental differences between our approach and that of the TW-dFC~\cite{Calamante2017}. Specifically, both approaches attempt to give a measure of FC into the WM using GM as a constraint. However, TW-dFC relies on two successive use of classical methods (e.g., tractoragphy and sliding-window FC), while our work introduces an entirely new concept where recovery of mediation by SC is incorporating the FC itself in a single integrated framework; i.e., the complete pattern of brain activity in GM is constraining the recovery of WM patterns. In addition, our method works at the single-frame level and does not require to choose a temporal window to render the approach dynamic.

\subsection*{New research avenues for structure-function studies}
GSP approaches as applied to functional neuroimaging rely on a \textit{graph shift operator} (typically the Laplacian or the adjacency matrix) that encodes the brain's anatomical information. Its eigendecomposition produces an orthogonal set of eigenvectors, termed \textit{eigenmodes}, some of which are reminiscent of well-established functional networks~\cite{Atasoy2016} and are useful in effectively estimating the strength of inter-hemispheric interactions in the brain~\cite{Robinson2016}. Moreover, efficient anatomically-informed decompositions of fMRI data using tractography-based structural connectome~\cite{Atasoy2016}, as well as topology encoding GM graphs~\cite{Behjat2015} have been proposed. Brain eigenmodes can be viewed as basic building blocks of increasing spatial variation along the structural brain graph, akin to sinusoids in classical Fourier analysis. Here, we extend the traditionally region-level eigenmodes (defined on a limited subset of GM nodes) to a whole-brain (GM and WM), voxel resolution setting (see Supplementary Fig.~S4), thus enabling to reconstruct structure-function networks at an unparalleled spatial resolution. As an increasing number of operations are generalized from classical signal processing to the graph setting~\cite{Shuman2013,Huang2018}, promising avenues arise to explore the many facets of brain structure and function.

We therefore foresee two direct avenues for future research. The first avenue aims on leveraging the proposed interpolated fMRI data. The interpolated volumes entail additional informative voxels that extend beyond the GM, which is in particular interesting in light of the limitations and skepticisms in interpreting WM BOLD data \cite{Logothetis,Gawryluk2014}. Previous works have demonstrated the possibility to capture functionally relevant information from the WM BOLD~\cite{Peer2017,Ding2018}, despite the well-established findings on the differences of hemodynamic responses in GM and WM~\cite{Fraser2012,Newton2019}. At its current form, we interpolate functional signals into the WM as solely constrained by the functional signals from the GM. Alternatively, one may rather consider combining signal interpolation with signal recovery of weak signals in the WM. In our current application, the goal is to observe WM pathways that support distributed patterns of FC in GM, and not on the functional organization of WM itself. The interpolated volumes can then be readily explored using existing tools for dynamic FC analyses, such as sliding-window correlation, ICA~\cite{Beckmann2005} and principal component analysis~\cite{Leonardi2013}, to probe functional brain dynamics at a much larger scale. 
The second foreseen research direction is to exploit the proposed anatomically-informed brain grid to implement novel GSP operations on fMRI data. High-resolution eigenmodes enable spectral graph-based analysis of task-based and resting-state fMRI data at an unprecedented level of detail, and across the whole brain, in contrast to the conventional region-wise analysis, which is typically also limited to the cortical layer. Anatomically-informed spectral graph decomposition of fMRI data using these high-resolution brain eigenmodes is anticipated to open up new perspectives on the brain's functional organization not only within GM but also WM. As a case in point, FC has often been associated to Euclidean distance~\cite{Markov2012,Alexander-bloch2013}; thanks to our framework, the notion of distance becomes more meaningfully defined in terms of the spectral representation of functional signals~\cite{Medaglia2018}, enabling a deepened understanding of the roles of short- and long-range connectivity in improving the efficiency of inter-areal brain communication. 

In conclusion, this study presents a new framework for studying structure-function relationship that has several key advantages. The interpolation of fMRI activity into the white matter enables observing key WM structures that link interacting GM regions.  We believe that the introduction of highly-resolved human brain eigenmodes can (i) shift the existing trend of constructing region-wise connectomes to that of voxel-vise connectomes and (ii) expand the use of GSP repertoire in the context of functional brain imaging. More importantly, we anticipate the proposed joint structure-function characterization to offer unprecedented benefits for the study of clinical populations hindered by complex mixes of structural and functional alterations, such as in patients with agenesis of the corpus callosum or virtual callosotomies~\cite{Park2008}.

\section*{Methods}

\subsection*{Data}
We used a total of 51 subjects obtained from the publicly available Human Connectome Project dataset (HCP 1200 release), WU-Minn Consortium. Preprocessed diffusion MRI, original functional MRI, and anatomical data are downloaded following a random subject selection scheme while maintaining a uniform demographic distribution (male and female, ages 22-50). MRI acquisition protocols of the HCP and preprocessing guidelines for diffusion MRI are fully described and discussed elsewhere~\cite{Glasser2013}. No further preprocessing was applied to the diffusion data. We then extracted the ODFs using DSI studio (http://dsi-studio.labsolver.org). On the other hand, original and unprocessed functional MRI sequences underwent a standard preprocessing protocol~\cite{VanDijk2009}. 

\subsection*{Building the voxel-wise brain grid}
Similar to classical brain graphs, we define our grid as $\mathcal{G}:= (\mathcal{V},\mathbf{A})$, where $\mathcal{V}=\{1,2,3,...,\mathit{N}\}$ is the set of $\mathit{N}$ nodes representing the brain voxels (i.e., $N = 700-900$ thousand), and $\mathbf{A} \in \mathit{N} \times \mathit{N} $ is an adjacency matrix that encodes the likelihood of water molecules to diffuse from their current position to neighboring voxels. We used an ODF-based weighting-scheme on a three-dimensional 26-neighborhood mesh. The topology of the brain grid is validated by defining the graph Laplacian matrix in its symmetric normalized form ($\mathbf{L} = \mathbf{I} - \mathbf{D}^{-\frac{1}{2}}\mathbf{A}\mathbf{D}^{-\frac{1}{2}}$), whose eigendecomposition leads to a complete set of orthonormal eigenvectors that span the graph spectral domain with their corresponding real, non-negative eigenvalues. The top 11 eigenfunctions corresponding to the lowest spatial variation are provided in Supplementary Fig.~S4.

\subsection*{Graph signal recovery formulation}
We consider a cost function that includes a least squares data-fitting term equal to the residual sum of squares (RSS), and an L2 regularization term that takes into account smoothness with respect to the brain grid, given by
\begin{equation}
\tilde{x} = \argmin_x  \left\|\mathbf{y}-\mathbf{Bx}\right\|^2 + \lambda\mathbf{x}^T\mathbf{L}\mathbf{x},
\label{gsr}
\end{equation}
where $\mathbf{y}$ is a vector of length $N$ containing initial BOLD values within GM nodes and zero elsewhere, $\mathbf{B}$ is an indicator matrix that selects the GM nodes, and $\lambda$ is the regularization parameter. $\mathbf{L}$ is the graph Laplacian, which encodes the voxel-level anatomical structure. The cost function balances between (1) minimizing the RSS and retaining the original GM signals, and (2) imposing smoothness with respect to the structure of the graph. The balance is dictated by the optimal $\lambda$ (see Supplementary Fig.~S5 for details on deriving an optimum value). The solution to Equation~\ref{gsr} gives the interpolated volume for one time-point. The interpolation is done repeatedly for each volume in the functional data.



\bibliographystyle{naturemag}
\bibliography{sample}


\section*{Acknowledgements}
We thank T. Bolton for his assistance and valuable comments. This work was supported by the Swiss National Science Foundation under the Project Grant 205321-163376.
\section*{Author Contributions}
 A.T., D.A., and H.B. designed the construction of the voxel-wise brain grid, A.T. and D.V.D.V. designed and performed research. A.T and D.V.D.V. wrote the manuscript.
 \section*{Competing Interests} 
 The authors declare that they have no competing financial interests.
\section*{Correspondence}
Correspondence and requests for materials should be addressed to A.T.~(email: anjali.tarun@epfl.ch).

\end{document}